\documentclass[useAMS,usenatbib]{mn2e}
\usepackage{psfrag,graphicx}
\usepackage{color}
\usepackage{txfonts}
\usepackage{breqn}
\usepackage[T1]{fontenc}
\usepackage{ae,aecompl}

\title[Black hole growth]
{Early growth of typical high redshift black holes seeded by direct collapse}
 \author[Latif  et al.]
{M. A. Latif\thanks{Corresponding author: latifne@gmail.com}$^{1,2}$,
Marta Volonteri$^{2}$,
John H. Wise$^{3}$\\
$^1$Department of Physics, COMSATS Institute of Information Technology, Park Road, 44000, Islamabad, Pakistan \\
$^2$Institut d'Astrophysique de Paris, Sorbonne Universités, UPMC Univ Paris 06 et CNRS, UMR 7095, F-75014, Paris, France \\
$^3$Center for Relativistic Astrophysics, Georgia Institute of Technology, 837 State Street, Atlanta, GA 30332, USA  \\
}
\date{}

\def\LaTeX{L\kern-.36em\raise.3ex\hbox{a}\kern-.15em
      T\kern-.1667em\lower.7ex\hbox{E}\kern-.125emX}

\begin{document}

\bibliographystyle{mn2e}

\label{firstpage}

\maketitle

\def\na{NewA}%
          % New~Astronomy
\def\aj{AJ}%
          % Astronomical Journal
\def\araa{ARA\&A}%
          % Annual Review of Astron and Astrophys
\def\apj{ApJ}%
          % Astrophysical Journal
\def\apjl{ApJ}%
          % Astrophysical Journal, Letters
\def\jcap{JCAP}

\def\apjs{ApJS}%
          % Astrophysical Journal, Supplement
\def\ao{Appl.~Opt.}%
          % Applied Optics
\def\apss{Ap\&SS}%
          % Astrophysics and Space Science
\def\aap{A\&A}%
          % Astronomy and Astrophysics
\def\aapr{A\&A~Rev.}%
          % Astronomy and Astrophysics Reviews
\def\aaps{A\&AS}%
          % Astronomy and Astrophysics, Supplement
\def\azh{AZh}%
          % Astronomicheskii Zhurnal
\def\baas{BAAS}%
          % Bulletin of the AAS
\def\jrasc{JRASC}%
          % Journal of the RAS of Canada
\def\memras{MmRAS}%
          % Memoirs of the RAS
\def\mnras{MNRAS}%
          % Monthly Notices of the RAS
\def\pra{Phys.~Rev.~A}%
          % Physical Review A: General Physics
\def\prb{Phys.~Rev.~B}%
          % Physical Review B: Solid State
\def\prc{Phys.~Rev.~C}%
          % Physical Review C
\def\prd{Phys.~Rev.~D}%
          % Physical Review D
\def\pre{Phys.~Rev.~E}%
          % Physical Review E
\def\prl{Phys.~Rev.~Lett.}%
          % Physical Review Letters
\def\pasp{PASP}%
          % Publications of the ASP
\def\pasj{PASJ}%
          % Publications of the ASJ
\def\qjras{QJRAS}%
          % Quarterly Journal of the RAS
\def\skytel{S\&T}%
          % Sky and Telescope
\def\solphys{Sol.~Phys.}%
          % Solar Physics

          % Solar Physics
\def\sovast{Soviet~Ast.}%
          % Soviet Astronomy
\def\ssr{Space~Sci.~Rev.}%
          % Space Science Reviews
\def\zap{ZAp}%
          % Zeitschrift fuer Astrophysik
\def\nat{Nature}%
          % Nature
\def\iaucirc{IAU~Circ.}%
          % IAU Cirulars
\def\aplett{Astrophys.~Lett.}%
          % Astrophysics Letters
\def\apspr{Astrophys.~Space~Phys.~Res.}%
          % Astrophysics Space Physics Research
\def\bain{Bull.~Astron.~Inst.~Netherlands}%
          % Bulletin Astronomical Institute of the Netherlands
\def\fcp{Fund.~Cosmic~Phys.}%
          % Fundamental Cosmic Physics
\def\gca{Geochim.~Cosmochim.~Acta}%
          % Geochimica Cosmochimica Acta
\def\grl{Geophys.~Res.~Lett.}%
          % Geophysics Research Letters
\def\jcp{J.~Chem.~Phys.}%
          % Journal of Chemical Physics
\def\jgr{J.~Geophys.~Res.}%
          % Journal of Geophysics Research
\def\jqsrt{J.~Quant.~Spec.~Radiat.~Transf.}%
          % Journal of Quantitiative Spectroscopy and Radiative Trasfer
\def\memsai{Mem.~Soc.~Astron.~Italiana}%
          % Mem. Societa Astronomica Italiana
\def\nphysa{Nucl.~Phys.~A}%
          % Nuclear Physics A
\def\physrep{Phys.~Rep.}%
          % Physics Reports
\def\physscr{Phys.~Scr}%
          % Physica Scripta
\def\planss{Planet.~Space~Sci.}%
          % Planetary Space Science
\def\procspie{Proc.~SPIE}%
          % Proceedings of the SPIE
\def\pasa{Pub.~Astro.~Soc. ~Aurstraila}%
% 

% \newcommand{\ch}[1]{\textcolor{red}{\textbf{#1}}}
% \newcommand{\newln}{\\&\quad\quad{}}
% \date{today}

 \begin{abstract}
 {Understanding the  growth of high redshift massive black holes (MBHs) is a problem of great  astrophysical interest.  The most luminous quasars at $z>6$ are frequently observed but they represent only the tip of the iceberg as the majority of the low luminosity AGN population remains undetected. In the present study, we perform a radiation hydrodynamics cosmological simulation to study the growth of ``normal" black holes  in the high redshift universe. In our simulation we  model the formation of Pop III and Pop II stars along with their chemical, mechanical and radiative feedback.  We consider  both UV and X-ray emission  from an accreting BH  to simulate its radiative feedback.  The selected halo has a mass of $\rm 3 \times 10^{10}~M_{\odot}$ at $z=7.5$ and we turn on radiative feedback from  a  MBH seed of $\rm 10^5~M_{\odot}$ along with in-situ star formation  at $z=12$ when the halo mass reaches well above the atomic cooling limit. We find that the MBH accretes only about 2200 $\rm M_{\odot}$ during 320 Myr and the average mass accretion onto the MBH is a few times $\rm 10^{-6}~M_{\odot}/yr$. Our results suggest that the stunted growth of MBH is a consequence of supernovae in tandem with  MBH feedback which drive large outflows and evacuate the gas from MBH vicinity.  This may explain why a population of low luminosity AGN has not been detected so-far at $z>6$; the large contrast between the star formation rate and the MBH accretion rate may make then hard to detect even in upcoming deep surveys.} 
 \end{abstract}

% conclusion 
\begin{keywords}
methods: numerical -- cosmology: theory -- early Universe -- high redshift black holes-- black holes physics-galaxies: formation
\end{keywords}

\section{Introduction} \label{sec:intro}
 Supermassive black holes (SMBHs) of millions to billions of solar masses are common at the centers of present day galaxies and may have co-evolved with their hosts (see \cite{Kormendy2013} and references therein). On the other hand, high redshift surveys have detected dozens of quasars at $z \geq 6$  powered by the accretion onto  SMBHs of  a few billion solar masses within the first Gyr after the Big Bang \citep{Fan2003,Willott2007,Jiang2009,MOrtlock2011,Banados2014,Venemans2015,Wu2015,Banados2017}.  These quasars represent  the tip of the iceberg as many low luminosity active galactic nuclei (AGN) may have gone undetected in present surveys due to current observational constraints.  Therefore, the population and growth of BHs remain elusive at earlier cosmic epochs. Understanding their formation and  growth mechanisms are questions of a prime astrophysical interest.
 
 Seeds of SMBHs may have formed at high redshift and  evolved via accretion or merging to reach a billion solar masses for the most extreme cases. The main pathways for SMBH formation are a collapse of a massive star, run-away collisions in stellar clusters and a direct collapse of gas cloud into a  massive BH, see a recent  review on this topic by \cite{Latif2016PASA}. The mass of seed BH depends on the formation channel and varies from $\rm 10 ~to~10^5~M_{\odot}$ \citep{Volonteri2010,Haiman2013}.  Large scale cosmological simulations exploring the growth of BHs commonly employ thermal feedback without proper radiation transfer calculations and  use Bondi prescription with a resolution of about a kpc \citep{Booth2009,Sijacki2015,DiMatteo2016} except \cite{Dubois2012} with a resolution of about 10 pc. They may overestimate the mass accretion onto  a MBH by not resolving the Bondi radius \citep{Gaspari2013,Gaspari17A,Negri2017,Johnson2016PASA}. Numerical experiments exploring the  growth of a stellar mass BH including a detailed treatment for its feedback show that gas in the surrounding  gets photo-evaporated and leads to its stunted growth shortly after its formation \citep{Johnson2007,Milos2009,Alvarez2009,Park2011}. Similarly, \cite{Jeon2012} and \cite{Jeon2014} found that BH feedback strongly suppresses its growth in minihalos and  also influences the star formation in the host galaxies. These studies  either employed idealized initial conditions or  focused on the growth of low mass BHs in the first minihalos forming at $z=10-30$, using cosmological simulations. 
 
 \cite{Johnson2011} performed  a 3D smoothed particle hydrodynamical simulation with  a ray tracing algorithm to model radiative feedback from a  MBH  in an atomic cooling halo. They found that accretion rate onto the BH drops down to $\rm 10^{-5}~M_{\odot}/yr$ in about one  Myr due to the photo-heating and radiation pressure.  However, these simulations were performed for primordial gas composition collapsing isothermally with no in-situ star formation.  The impact of X-ray feedback  from  a MBH seed on its growth was explored via cosmological radiation transfer simulations by  employing the adaptive mesh refinement (AMR) technique in a halo of $\rm \sim 10^8~M_{\odot}$  \citep{Aykutalp2013, Aykutalp2014}. They found that X-ray feedback from BH  self-regulates its growth and also induces star formation.  Recently,  \cite{Smidt2017} performed radiative transfer cosmological simulations to explore the role of X-ray feedback on its growth  assuming a single frequency bin  at 1 keV in the context of the most luminous high redshift quasars. They found that dense cold accretion flows feed  MBH which grows up to billion solar masses  by  $z=7$ and may explain the existence of $z> 6$ quasars.
  
In this study, we focus on the growth of ``normal" BHs, not the extreme population such as the brightest quasars at $z > 6$  in the early universe. To accomplish this goal, we perform an AMR cosmological simulation coupled with radiative transfer algorithm \textsc{moray}  \citep{Wise2011}.  We model  feedback from a MBH by including  both UV and X-ray (11.2 eV - 1.1 keV) emission from  an accreting BH of $\rm 10^5~M_{\odot}$.  We also follow the formation of Population III (Pop III) and Population II (Pop II) stars along with their radiative, chemical and mechanical feedback.  For star formation, we employ recipes  for Pop III and Pop II from \cite{Wise2012} which allows transition to second generation of stars for $\rm Z/Z_{\odot}> 10^{-4}$. Pushing the state of the art in this field, this work provides better insight about the growth of normal black holes at high redshift ($z >6$). 

This article is organised in the following way. In section 2, we describe in detail the simulation setup and  prescriptions for  feedback from MBH/stars as well as recipes for star formation. We present our main results in section 3 and confer our conclusions in section 4.

\section{Numerical methods}
We conduct the simulation using the public version of cosmological hydrodynamical  code \textsc{enzo} \citep{Enzocode2014}.  It is  an AMR grid based code which makes use of the  message passing interface (MPI) to run on parallel systems.  The equations for hydrodynamics are solved with the piece-wise parabolic method (PPM) and we employed the HLLC Riemann solver to capture strong shock and rarefaction waves.  The code makes use of  a particle-mesh based N-body solver to compute dark matter (DM) dynamics. A multi-grid Poisson solver is used for self-gravity calculations.

We start our simulation at $z=150$ with cosmological initial conditions generated from the MUSIC package \citep{Hahn2011} and use the latest Planck  data  with $\Omega_{m}=0.3089$, $\Omega_{\Lambda}=0.6911$, $\rm H_{0}=0.6774$ to generate initial conditions \citep{Planck2016}. Our  periodic box has a size of comoving 8~Mpc/h on each side, we select the most massive halo forming in our computational volume and place it at the center of the simulation box.  We employ nested grid initial conditions with a root grid resolution of $\rm256^3$ cells, an equal number of DM particles and two additional nested grids with $\rm 256^3$ cells, resulting in an effective resolution of $\rm 1024^3$.  This setup provides us an effective DM resolution of $\rm \sim 53,000 ~M_{\odot}$. We further employ 8 additional  dynamic refinement levels during the course of simulation which yields a maximal physical resolution of  3.6 pc at $z=11$.  Our resolution criteria is based on  the baryonic overdensity of 4, the particle mass resolution of $\rho_{DM} \Delta x^3 r^{\ell \alpha}$  where $\rho_{DM}$ is the dark matter density, $\Delta x$ is the root grid cell size, $r = 2$ is the refinement factor, $\ell$ is the refinement level, and $\alpha = -0.3$ makes the refinement super-Lagrangian and a fixed Jeans resolution of at least four cells, similar to \cite{Latif2016}.  The simulated halo has a mass of $\rm 3 \times 10^{10}~M_{\odot}$ and $\rm 6.2 \times 10^{10}~M_{\odot}$ at $ z=7.5$ and $z=6$, respectively. The virial radius of the halo at $z=7.5$ is 21.5 kpc. We follow its formation from $z=14$ down to $z=7.5$,  turn on star formation and feedback from the MBH at $z=12$ when the halo mass is $\rm \sim 10^9~M_{\odot}$ and it is resolved by 20,000 DM particles.  We did not allow  star formation in the halo before $z=12$ assuming that one of its progenitors  formed a MBH of $\rm 10^5~M_{\odot}$ via direct collapse \citep{Latif2013d,Johnson2013b}. We mention the consequences of such delayed star formation in the discussion section.
%DM particle density higher than  0.0625 times  $\rho_{DM}r^{\ell \alpha}$  where $\rho_{DM}$ is the dark matter density, r = 2 is the refinement factor, $\ell$ is the refinement level, and $\alpha = -0.3$ makes the refinement super-Lagrangian

We employ a non-equilibirum time dependent chemical model which solves rate equations of nine primordial species $\rm H,~H^+,~ H^-, ~He,~ He^+, ~He^{++},~ H_2, ~H_2^+, ~e^-$.  Our chemical model is mainly based on \cite{Anninos1997} and \cite{Abel97}.  It includes  cooling due to the collisional excitation, collisional ionization,  radiative recombination,  Bremstrahling radiation, $\rm H_2$ cooling and Compton heating/cooling. In addition to primordial cooling, we also include metallicity dependent  metal lines (C, N, O, Si etc) cooling  from \cite{GloverJappsen2007} which operates in the regime $100-10^4$ K, above $\rm 10^4$ K, the cooling function of \cite{Sutherland1993}  is employed.  We only consider metal line cooling  from metals produced  by Pop III stars and ignore the effect of cooling by metals produced by Pop II stars,  doing so we may underestimate the stellar mass by a factor of two, for further details see  \cite{Wise2012a}.  We assume a constant $\rm H_2$ dissociating flux of strength $\rm J_{21}=500$, where $\rm J_{21}$ is in units of  $\rm 10^{-21} erg/s/cm^2/Hz$. Our chemistry solver is coupled with the radiative transfer module \textsc{moray} to take into account heating from photoionization and secondary ionization heating \citep{Shull1985} from X-rays emitted by a MBH. 

\subsection{Star formation}
Our star formation criteria are primarily based on \cite{Wise2008A} and \cite{Wise2012} for  Pop III and  Pop II stars, respectively, which we briefly summarise here. We  model  the formation of both Pop III and Pop II stars and distinguish them based on the metallicity of gas cloud.  Pop II stars in our simulation are formed for $\rm Z/Z_{\odot} > 10^{-4} $, otherwise Pop III stars form. A Pop III star particle is allowed to form when a cell has an over density of $5 \times 10^5$ ($\rm 10^3 ~cm^{-3}$ at $z=10$), a molecular hydrogen ($\rm H_2$) fraction of $ \rm \geq 5 \times 10^{-4}$ and the flow is convergent ($\rm \nabla  \cdot v_{gas} <0 $). Each Pop III star particle represents a single star and its mass is randomly sampled from the initial mass function (IMF) with mass range between 1-300 $\rm M_{\odot}$ with the following shape,

\begin{equation}
 f ( $\rm log M$) dM = M^{-1.3} \exp \left[  -  \left(\frac{M_{char}}{M} \right)^{1.6} \right] dM
\label{eq0}
\end{equation}
where $M_{char}$ is the characteristic mass and is set to 100 $\rm M_{\odot}$. The chosen IMF behaves as a Salpeter IMF above $\rm M_{char}$ and has an exponential cutoff below it.  The Pop II star formation criterion is similar to Pop III but the requirement for minimum $\rm H_2$ is removed as  gas in the presence of metals can cool down even in the presence of a UV background. Pop III star particles are restricted to form in cells with temperature $\rm < 1000$ K. Contrary to a Pop III star particle,  a Pop II star particle represents a star cluster with a minimum mass of $\rm 1000 ~M_{\odot}$ with a Salpeter IMF. Pop II star particles that are born with $\rm < 1000 ~M_{\odot}$ continue to accrete gas until they cross this threshold. For cells meeting this criteria, 7 \% of the cold gas ($\rm T< 1000 ~K$) within a sphere over a dynamical timescale of about 3 Myr ($\rho_{av} \sim  1000 ~ \mu~cm^{-3}$) is converted into stars. Both newly forming Pop III and Pop II star particles are merged within a radius of 10 pc in a single timestep.

\subsection{Stellar feedback}
Star particles are treated as point sources of radiation, and the radiative feedback from stars is modelled using the adaptive ray tracing scheme based on HEALPix framework \citep{AbelWand2002,Wise2011} which is self-consistently coupled with the hydrodynamics. Moreover, star particles emit  ionising radiation with a monochromatic spectrum. The $\rm H_2$ dissociating radiation (LW) from radiation sources is modelled with an inverse-square profile around them, optically thin limit. The mass dependent hydrogen ionizing and LW luminosities of Pop III stars are taken from \cite{Schaerer2002}. For  Pop III stars, we take a monochromatic spectrum with an energy of 29.6 eV suitable for stars with  surface temperatures of $\rm 10^5$ K, while for Pop II stars, we take a fixed spectrum of 21.6 eV appropriate for low metallicity stars \citep{Schaerer2003}. Pop II stars  emit 6000 hydrogen ionizing photons per  stellar baryon for 20 Myr, or equivalently $\rm 2.4~\times~10^{47} ~photons~s^{-1}~M_{\odot}^{-1}$.

Depending upon their mass,  Pop III stars end their lives either as a supernovae (SNe) or collapse into a BH. We model SN energy and also X-ray feedback from a stellar mass black holes via ray tracing module mentioned above. Pop III stars with masses between 11-40 $\rm M_{\odot}$ and 140-260 $\rm M_{\odot}$ die as Type II SNe and  pair-instability SNe (PISNe), respectively \citep{Heger2003}. While Pop III stars with mass between 40-140 $\rm M_{\odot}$ and above 260 $\rm M_{\odot}$  are expected to directly collapse into a BH  \citep{Heger2002, Heger2003}.  The typical energy for type II SNe  from 11-20 $\rm M_{\odot}$ stars is $\rm 10^{51}~erg~s^{-1}$ and  for 20-40 $\rm M_{\odot}$ stars energy is taken  from \cite{Nomoto2006} while for  PISNe is $\rm 10^{51}-10^{53}~erg~s^{-1}$ \citep{Heger2002}. Pop II star clusters  generate  $\rm 6.8 \times 10^{48}~erg~s^{-1}M_{\odot}^{-1}$ from SNe  4 Myr  after their formation, which is distributed in a sphere of radius 10 pc. 

\subsection{Massive black hole feedback}
We insert a massive black hole (MBH)  seed of $\rm 10^5~M_{\odot}$ at the halo center at $z=12$  possibly formed via direct collapse \citep{Latif2013d} and treat it as collionless sink particle that can grow via accretion.  For the spectral energy distribution (SED) of a MBH, we compute multicomponent spectrum which includes both the multi-color black body and a power law component $\propto E^{-\alpha}$ with $\alpha =1$ \citep{Shakura73,Sazonov2004}. We model both UV and X-ray feedback from an accreting  MBH  by considering the energy range from 11.2 eV-1100 eV. Four energy bins  at 13.6 eV, 24.6 eV, 54.4 eV and 1 keV are used when generating the photons. We treat  the MBH particle as a radiation source and perform 3D radiative transfer calculations to model its radiative feedback by making use of radiative transfer module \textsc{moray} that couples it with hydrodynamics. We have extended the  implementation of a radiative feedback from a MBH by adding more energy bins, originally from \cite{Kim2011}. The luminosity of an accreting black hole can be estimated as:
\begin{equation}
L_{BH} = \epsilon_{r} \dot{M}_{BH} c^2
\label{eq1}
\end{equation}
where $\epsilon_{r}$ is the radiative efficiency, assumed to be 0.1, $\dot{M}_{BH}$ is the MBH accretion rate and c is the speed of light.  We estimate the total number of ionizing photons from MBH as $\rm L _{BH} \times dt_{ph}/E_{av}$ where $dt_{ph}$ is the radiation transport timestep and $E_{av}$ is the average energy in the given frequency bin. These photons interact with the surrounding gas by photo-ionizing hydrogen and helium atoms,  even causing secondary ionizations via suprathermal electrons, and  photo-heating the gas. For a detailed description about these processes, we refer the reader to \cite{Kim2011}. We model the gas accretion onto the MBH following the prescription of  \cite{Kim2011} and estimate $\dot{M}_{BH}$ using the Eddington limited Bondi-Hoyle equation as follows:

\begin{equation}
\dot{M}_{BH} = $\rm min$ (\dot{M}_B, \dot{M}_{Edd})  =  $\rm min$ \left( \frac{4 \pi G^2  M_{BH}^2 \rho_{B}}{c_s^3},  \frac{4 \pi G M_{BH} m_p}{\epsilon_r \sigma_T c} \right)
\label{eq2}
\end{equation}
where $G$ is the gravitational constant, $M_{BH}$ the black hole mass, $\rho_{B}$ is the density at the Bondi radius, $c_s$ is the sound speed,  $m_p$ is the proton mass, $\sigma_T$ is the Thomson scattering cross-section. The Bondi accretion radius is given by:
\begin{equation}
R_{B} = \frac{2 G M_{BH}}{c_S^2} .
\label{eq3}
\end{equation}
A portion of gas with mass $\dot{M}_{BH} \Delta t$ inside the Bondi radius is accreted onto a MBH and is uniformly subtracted from grid cells within the Bondi sphere.  The maximum resolution in our simulation is 3.6 pc, which is close to the Bondi radius for $\rm 10^5~M_{\odot}$ black holes sitting in $\rm T \sim 8000$ K gas (i.e. 8.6 pc). Therefore, the Bondi radius is resolved at most times, when the gas temperature is  allowing us to probe the gas dynamics onto accreting MBH.

%We would like to mention here that  the Bondi radius in our simulation is resolved  at most  times which is imperative to probe the gas dynamics onto accreting MBH.

\begin{figure*}
\includegraphics[scale=0.5]{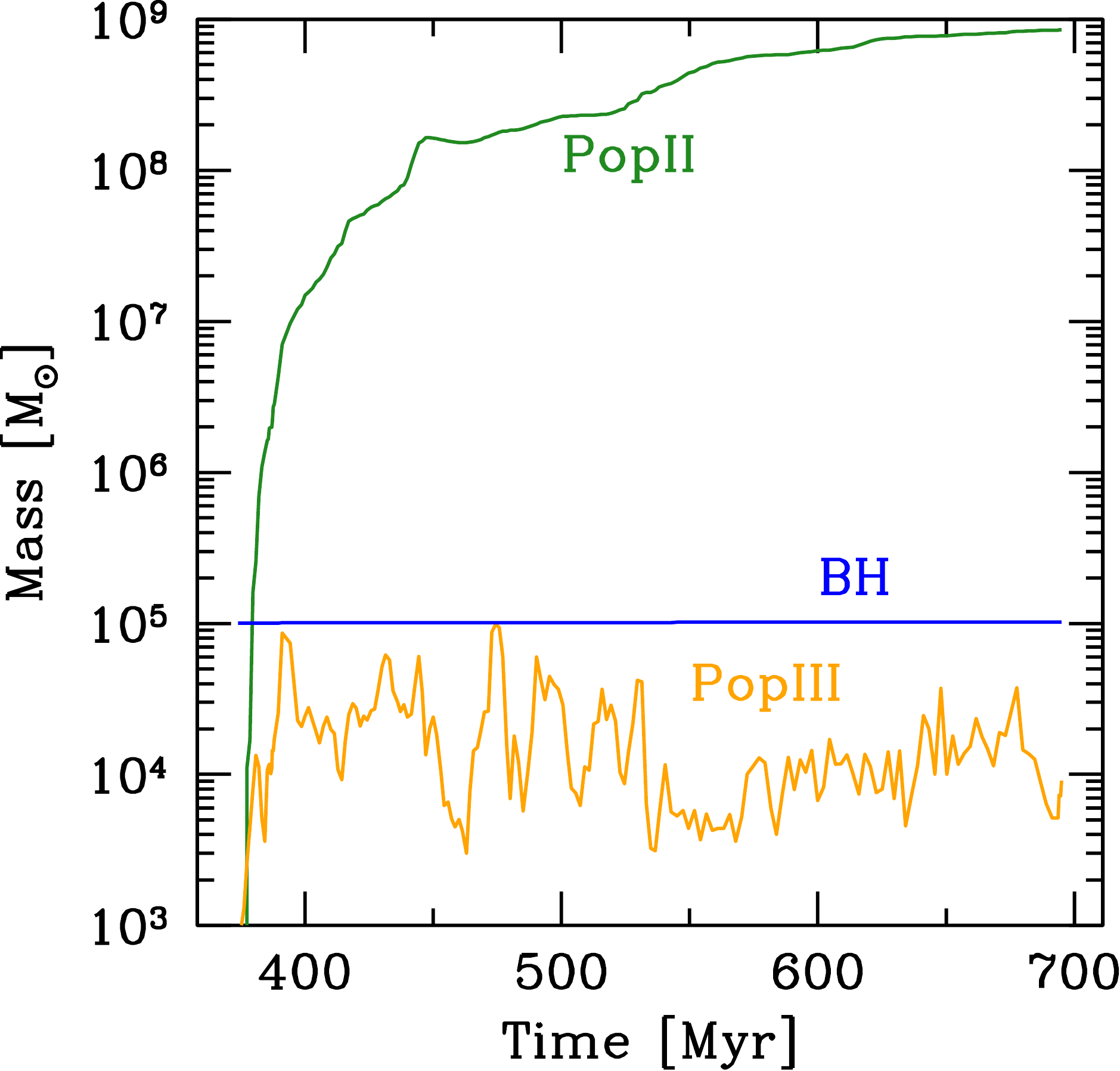} 
\caption{Time evolution of stellar and BH masses. The green line shows mass in Pop III stars,  orange line represent Pop II stellar mass and blue line shows the MBH mass.  Time is measured in Myr after the Big Bang. }
\label{fig1}
\end{figure*}

 \section{Results}
In this section we present results from a 3D cosmological radiation hydrodynamical simulation performed to explore the growth of ``normal" BHs within the first Gyr after the Big Bang. The simulated halo has a mass of $\rm 3 \times 10^{10}~M_{\odot}$ at $z=7.5$ and we follow its formation from $z=14$ down to $z=7.5$. We insert a MBH particle of $\rm 10^5~M_{\odot}$ at $z=12$ when halo mass reaches $\rm \sim 10^9~M_{\odot}$, well above the atomic cooling limit, and  turn on radiation feedback from the MBH.  The merger history of the simulated halo shows that it has gone through three major mergers (mass ratio greater than 1:2) at cosmic times of 425 Myr, 586 and 645 Myr,  respectively, and many minor mergers. Our simulation does not resolve star formation in halos below $\rm \sim 10^9~M_{\odot}$ at $z >12$ and we discuss its implications in the caveats section.
%This cosmological radiative transfer simulation includes Pop III and Pop II  star formation along with their chemical, mechanical and radiative feedback as well as  both UV and X-ray feedback from an active MBH.
\begin{figure*}
\includegraphics[scale=0.27]{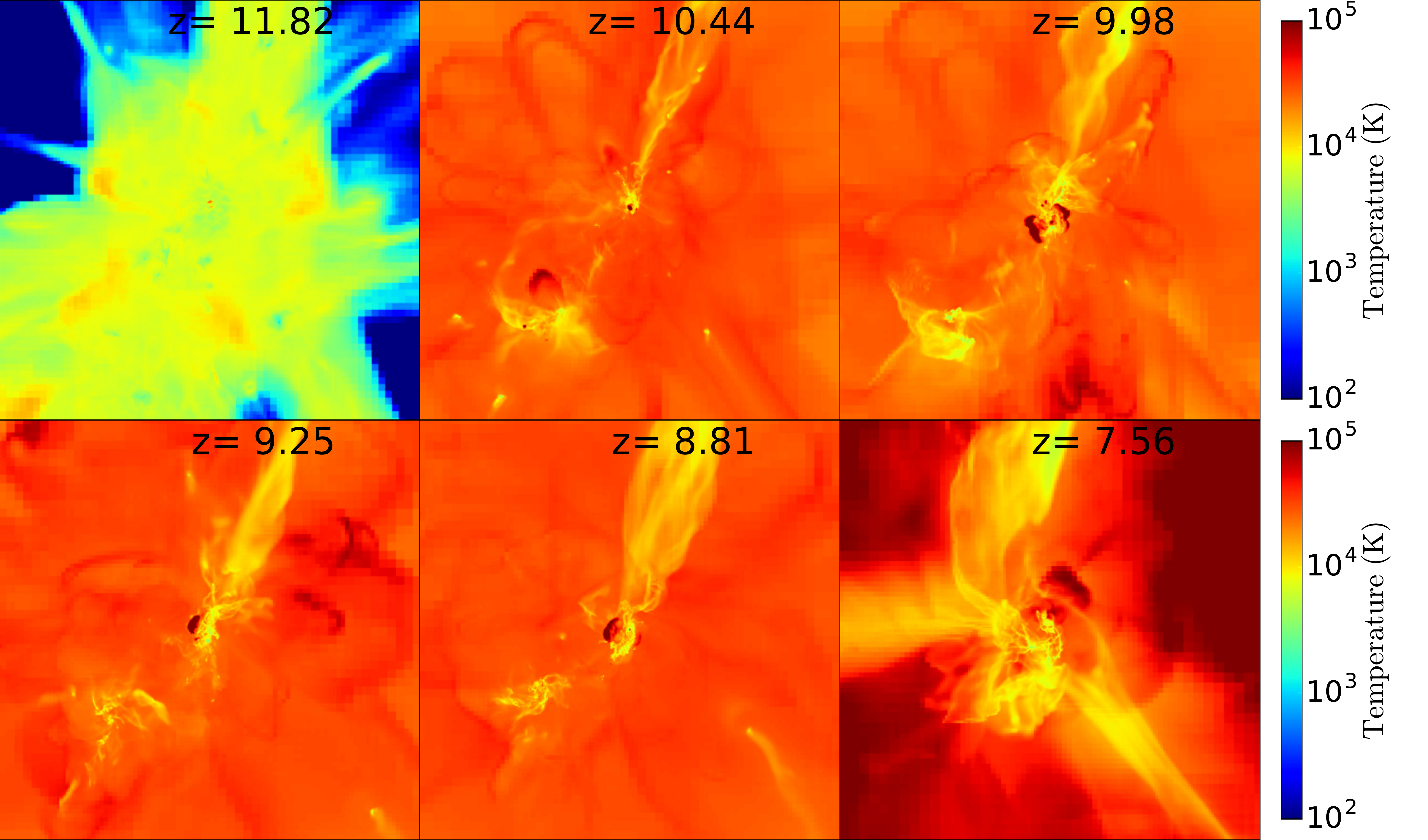}
\caption{Projections of density-weighted  temperature  along the y-axis at various redshifts for the central 14 kpc region. The top left panel shows the moment when first SN goes off while other panels depict the epochs when large outflows are produced by the SNe in tandem with MBH feedback.}
\label{fig2}
\end{figure*}

\begin{figure*}
\includegraphics[scale=0.27]{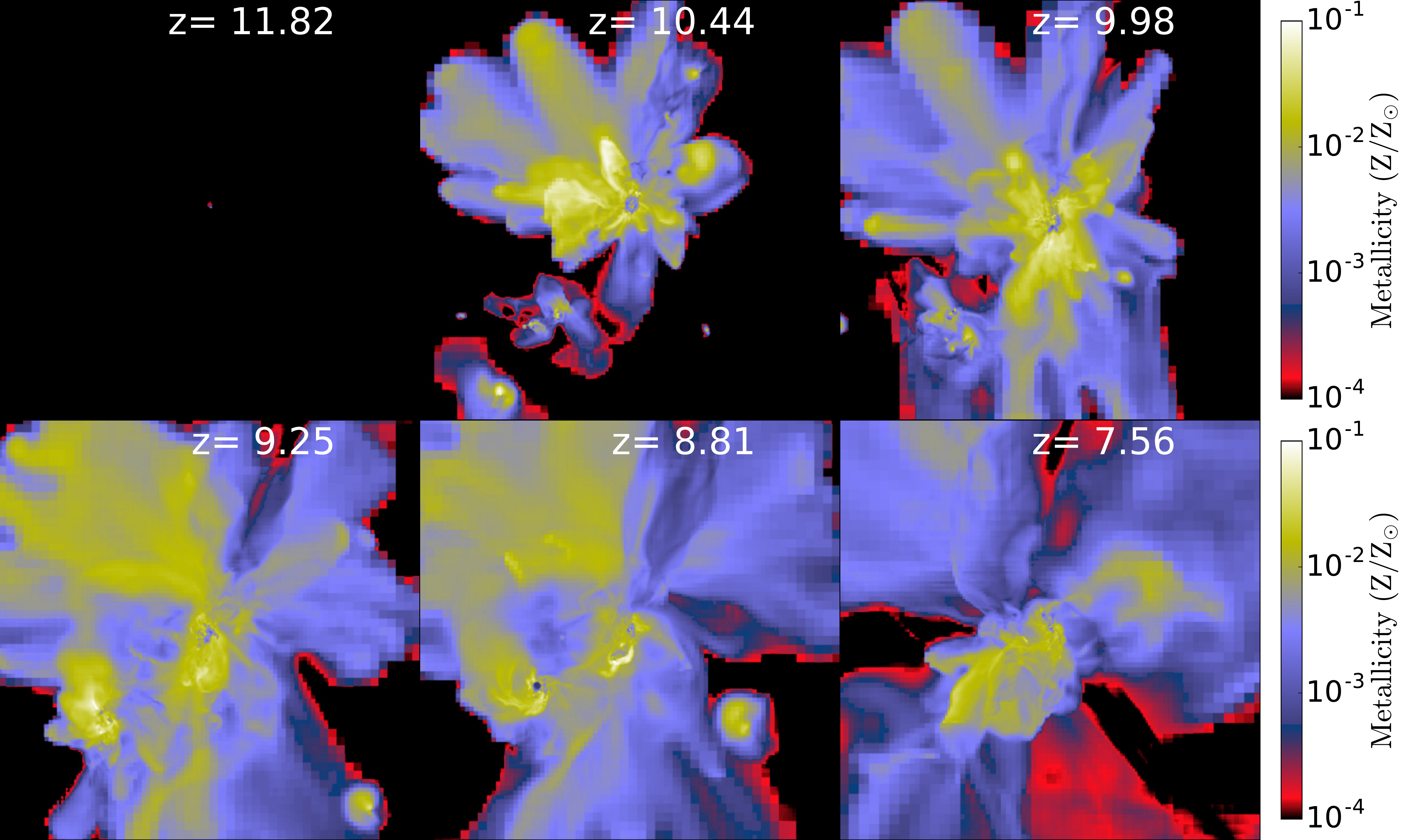}
\caption{ Projections of density-weighted metallicity originating from Pop III SNe along the y-axis at the same redshifts as figure \ref{fig2} for the central 14 kpc region. }
\label{fig3}
\end{figure*}

\begin{figure*}
\includegraphics[scale=0.27]{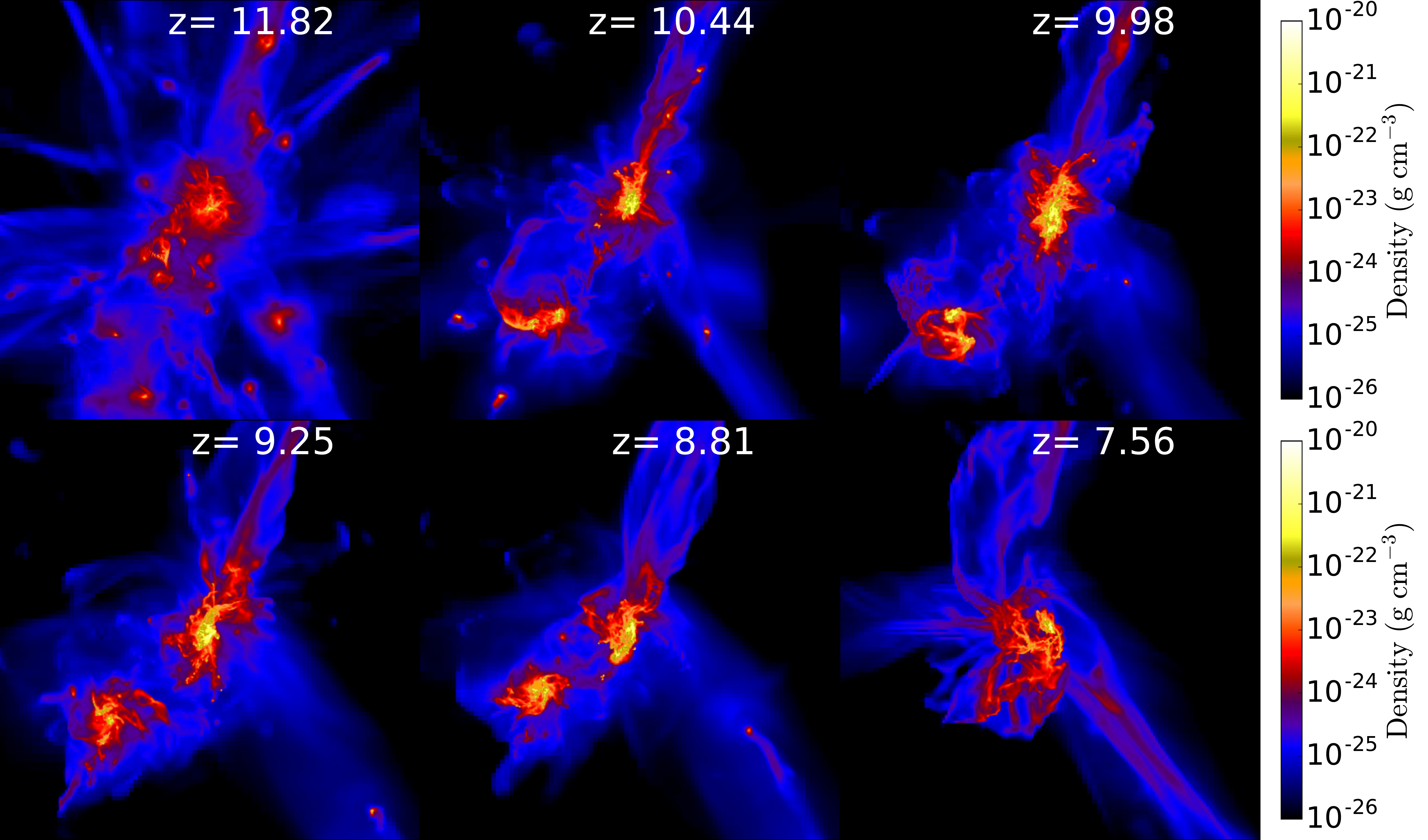} 
\caption{Projections of density-weighted gas density along the y-axis at the same redshifts as figure  \ref{fig2} for the central 14 kpc region. }
\label{fig4}
\end{figure*}

\subsection{Star formation}
In our simulation Pop III stars begin to form in  pristine $\rm H_2$ clouds self-shielded from the MBH radiation, 2-3 Myr after turning on the star formation which is restricted to the main halo. They explode as supernovae within a few Myr after their formation due to  high masses and enrich the interstellar medium with metals. The first Pop II star cluster forms 3 Myr  after the birth of Pop III stars, immediately after SNe. Both Pop III  and Pop II stars continue to form but Pop II stellar mass overtakes  the Pop III stellar mass  after 5 Myr.  The decline in  Pop III stellar mass by a factor of a few after 13 Myr  is due to the death of massive stars, and it  later on increases up to $\rm 10^5~M_{\odot}$. Pop III stars keep forming in metal free minihalos and at the same time they  continue to explode as SNe. Figure \ref{fig1} shows that the average Pop III stellar mass over the simulated time is  $\rm \sim 10^4 ~M_{\odot}$ while  wiggles on the Pop III stellar mass plot indicate the epochs of starburst and massive SNe going off. Peaks in the Pop III stellar mass correspond to the  mergers of  minihalos occurring at those epochs. Star formation mainly occurs in the primary halo and remains inhibited in the surrounding halos due to the feedback from primary halo.

A large number of SNe  generate  massive galaxy-wide outflows and particularly these outflows are observed at 450 Myr, 497 Myr, 560 Myr  and 640 Myr, where time is the age of the universe, as shown in Figure \ref{fig2}.  SNe along with AGN feedback heat the gas above $\rm 10^4$ K already at $z=10$. Highly anisotropic HII regions are formed at these epochs which propagate through the interstellar medium and heat the surrounding gas. By $z=7.5$, the HII region extends out to 20 kpc. These SNe keep enriching the halo with metals, and  the Pop II stellar mass increases over time due to the efficient metal cooling, culminating in $\rm \sim 10^9 ~M_{\odot}$ of Pop II stars by $z = 7.5$.  As soon as first Pop III star goes SNe, it enriches the medium with metallicity of $\rm 10^{-2} ~Z_{\odot}$ well above the critical metallicity and a Pop III star cluster forms, see the top left panel of Figure \ref{fig3}. The average metallicity enhancement from Pop III stars alone is about $10^{-2} ~Z_{\odot}$ and our estimates for the metallicity from Pop III stars are shown in Figure \ref{fig3}. Metal enrichment is very inhomogeneous and extends beyond the virial radius. Also, pockets of  metal poor ($\rm Z/Z_{\odot} \leq 10^{-4}$) gas exist down to $z=7.5$ in filaments.  

The density structure inside the halo is quite clumpy and turbulent which leads to chaotic accretion, also see \cite{Gaspari17}. The  state of gas density at various epochs is shown in Figure \ref{fig4} which depicts the occurrence of major merger between $z=8.8$ and z = 7.5. The peak density is about $\rm 10^4 ~cm^{-3}$ while the typical gas density is about $\rm 10-100 ~cm^{-3}$.  The overall density  structure in the halo becomes filamentary by the end of the simulation,  shown in Figure \ref{fig4}.  The average star formation rate (SFR) in our simulations for Pop III stars is $\rm 10^{-3}~M_{\odot}/yr$  while the average SFR for Pop II stars is a few  $\rm M_{\odot}/yr$ and peaks around $\rm 10 ~M_{\odot}/yr$. The Pop II SFR falls below $\rm 1 ~M_{\odot}/yr$ during the  strong outflows as they evacuate gas from the halo. Overall,  the average SFR is $\rm \sim 2~ M_{\odot}/yr$ and peaks in  SFR correspond to the mergers occurring at  those epochs.  Most of the SF occurs inside the primary halo. These results are comparable to the SFR main sequence at $z=6.5$  from the CANDELS survey \citep[see Fig.~\ref{fig2}]{Salmon2015}. The typical star formation rates observed in the CANDELS survey are between $1-100~M_{\odot}/yr$, with stellar masses of  a few times $\rm 10^8-10^{10}~M_{\odot}$ and z between $3.5-6.5$. Although, the simulated galaxy is at z=7.5, at slightly higher redshift, but its stellar mass and the SFR are close to the observed range at z=6.5. Such a comparison  confirms that the simulated galaxy is a typical,  not completely off the main sequence observed at $z>6$ and  agree with predictions from theoretical models  \citep{Behroozi2013,Dave2013,Somerville2012}.

\begin{figure*}
\includegraphics[scale=0.5]{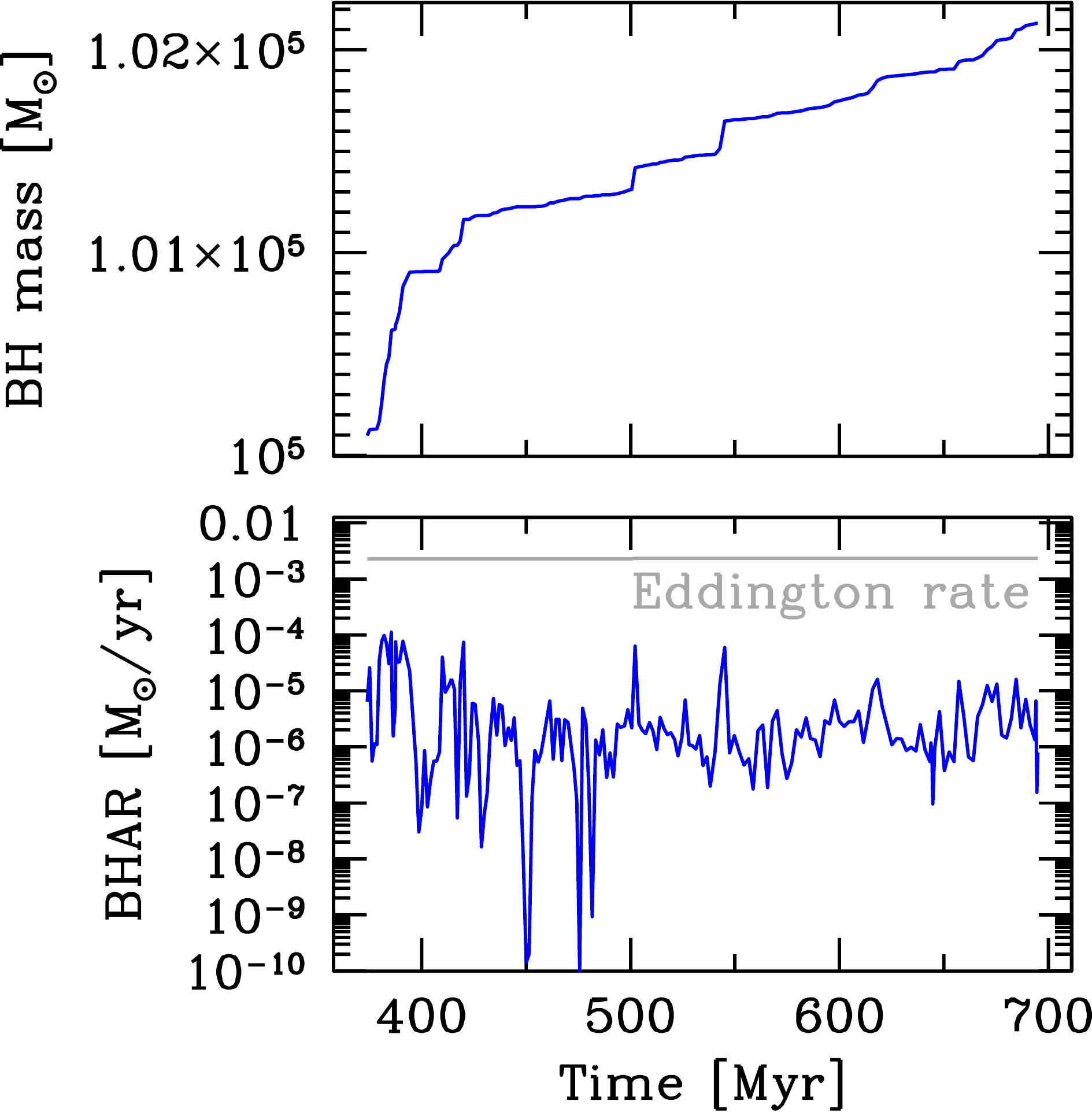}
\caption{The black hole mass evolution (top panel) and  the  mass accretion rate onto the MBH (bottom panel) over the cosmic time in Myr after the Big Bang, showing the lack of the growth of ``normal" BHs in the early universe. }
\label{fig5}
\end{figure*}

\begin{figure*}
\includegraphics[scale=0.5]{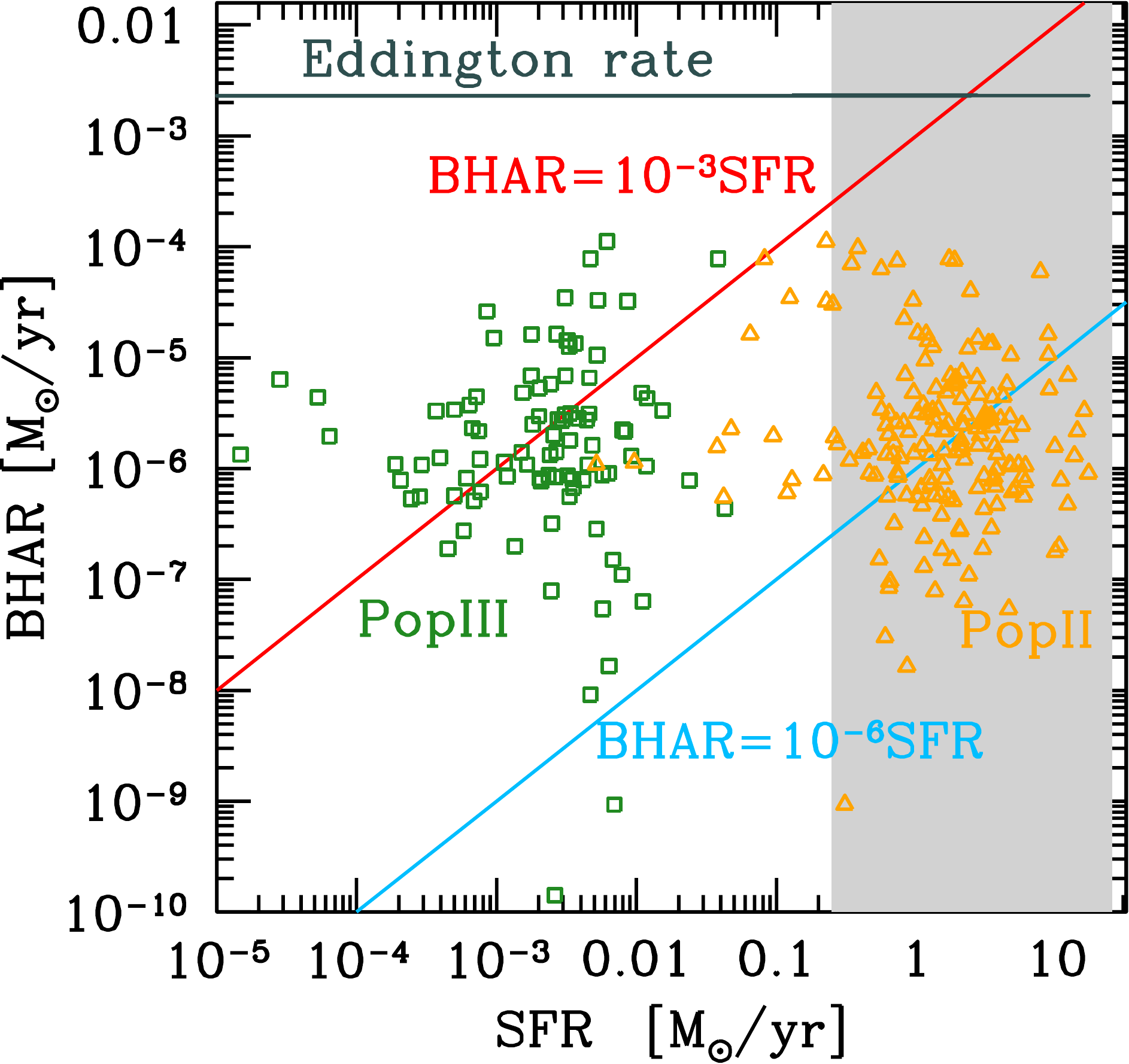} 
\caption{The MBH accretion rate against star formation rate is shown here. The green squares represent  Pop III SFRs while orange triangles correspond to Pop II SFRs.  They are plotted for all halo progenitors at all redshifts. The grey-shaded region shows the range of SFR found in Salmon et al. 2015. }
\label{fig6}
\end{figure*}

 \subsection{Black hole growth}
The  initial mass accretion rate  onto the MBH is $\rm \sim 10^{-5}~M_{\odot}/yr$  and continues to increase for about  first 3 Myr.  The X-ray and UV feedback from the MBH photo-dissociate molecular hydrogen and photo-ionize hydrogen and helium atoms in its surrounding.  Such energy deposition photo-evaporates the clumpy gas and forces it away from the MBH. The  luminosity from young stars exceeds the MBH luminosity within the central kpc region which further heats and expels the gas from the central potential. As a result the temperature of  gas rises  to $\rm \geq 10^4$  K, and the expanding HII region leaves behind a low density medium.  In the meantime the first Pop III SN occurs that further pushes the gas away from the MBH and generates an outflow.  Consequently, mass accretion  onto the MBH drops below $\rm 10^{-6}~M_{\odot}/yr$ as shown in Figure \ref{fig5}. In the aftermath of SN, metal cooling brings the gas temperature down to $\rm \leq 10^4$ K and increased accretion  onto the MBH starts again after a few Myr. Accretion onto the MBH continues until many SNe explode.  They in tandem with MBH feedback expel the gas from MBH neighborhood and drive large outflows.  On a few occasions  the MBH gets trapped inside SN remnants. Subsequently, the accretion rate drops significantly and the growth of the MBH  is halted for a few Myr.  As mentioned above, large outflows observed at about 400 Myr, 450 Myr,  480 Myr, 560 Myr and 640 Myr strongly influence the MBH growth and particularly their impact  is severe  when the stellar mass is less than $\rm 10^8~M_{\odot}$.

The mass accretion onto the MBH reaches brief maxima of $\rm 10^{-4}~M_{\odot}/yr$, but starbursts generate large  kpc-wide outflows and regulate the growth of MBH.  This trend has been observed particularly for the first 150 Myr of simulation similar to previous works \citep{Dubois2012,Habouzit2016b, Prieto2017}. The ratio of the MBH mass accretion rate  to the Eddington accretion rate, so-called  the Eddington ratio, increases up to a  $\rm  \sim 0.03$ in the first 20 Myr and  declines during the massive outflows generated by SNe. The  average Eddington ratio during the simulation is $\rm \sim10^{-3}$. In our simulation the MBH never accretes at the Eddington limit.  

The average accretion rate onto the MBH  is a few times $\rm 10^{-6}~M_{\odot}/yr$ and  remains highly intermittent throughout the simulation. In total about 2200 $\rm M_{\odot}$ are accreted onto the MBH during 320 Myr  out of which  about 1000 $\rm M_{\odot}$  is accreted during the first 40 Myr.  Over time, the halo and stellar mass increase as the  potential well gets deeper. Therefore, the average density in the halo increases  as well as  the average temperature due to the continuous heating of gas by SNe and AGN feedback.  Consequently, the growth of the MBH tends to be faster during the last 50 Myr.  In Figure \ref{fig6}, we plot the accretion rate onto the MBH  against  the SFR. The ratio between MBH accretion rate and SFR is $\rm \sim 10^{-3}$ for Pop III stars and  $\rm \sim 10^{-6}$ for Pop II stars; this confirms that the MBH grows much more slowly than its host galaxy, since SFR and  therefore stellar mass are dominated by Pop II stars. Unless MBH accretion becomes more consistent, the MBH-galaxy system will drift towards a smaller ratio between MBH and stellar mass, creating a MBH ``undermassive" with respect to its galaxy. The large ratio between SFR and MBH accretion rate would also make it difficult to detect the AGN, it's luminosity likely swamped by that of the stellar population \citep[see][]{Volonteri2017}.

%The dense cold streams of gas  with density $\rm \sim \geq 1~cm^{-3}$ flow into the halo center and feed MBH, see figures \ref{fig6} and \ref{fig7}.

Overall, the stunted growth of  MBH is a consequence of  SNe in tandem with BH feedback which expel  gas from the MBH surroundings and quench its growth. These findings are  consistent with previous works exploring the growth of BHs at early cosmic times \citep{Dubois2012,Dubois2015,Habouzit2016b, Prieto2017,DiMatteo2016,Alcazar17,Smidt2017,Biernacki17}.  They also found that MBH does not grow efficiently in high redshift galaxies when stellar mass is below $\rm 10^9~M_{\odot}$. These studies ignored the formation and feedback from Pop III stars which can be important during the early stages of galaxy formation.  By  following the formation and feedback from Pop III stars, we found that it makes the growth of MBH even more difficult as massive Pop III stars could produce PISNe which are more effective in ejecting the gas from central potential, being 10-100 times more energetic than Type II SNe. Consequently, in comparison with earlier findings, the mean accretion rate onto the MBH is about two orders of magnitude smaller. These differences arise from modeling the feedback from Pop III stars and inclusion of both UV and X-ray feedback from the MBH. Our results are more robust than previous estimates as we perform radiative transfer calculations to model the radiative feedback from both stars and the central MBH.  

Recently, it was found  from BLUETIDES simulations that MBH can grow up to a few times $\rm 10^8~M_{\odot}$ already at $z=8$ in halos sitting in low tidal fields \citep{DiMatteo2016}. It should be noted that the halo mass in our simulation is  factor of ten lower  compared to  \cite{DiMatteo2016} but  the spatial resolution is about 50 times higher. Moreover, they employed thermal feedback  to model  feedback from MBHs while we performed  radiative transfer calculations. MBH does not grow much in our case, most likely the differences arise from  the environment of the halo,  its merger history and  the sub-grid physics. The other works  exploring the growth of MBH had a spatial resolution of about few hundred pc \citep{Feng2014,Costa2014,Schaye2015,Feng2016} and used  thermal feedback from MBH.  Doing so, they may have overestimated the mass accretion onto MBH, see \cite{Negri2017} for a comparison of different resolutions and  consequences of variations in the implementation of thermal feedback recipes.

\section{Discussion and Conclusions}
We have performed  a cosmological radiation hydrodynamical simulation to explore the growth of ``normal" BHs in the first Gyr after the Big Bang. To accomplish this goal, we selected a halo of $\rm 3 \times 10^{10}~M_{\odot}$ at $z=7.5$.  Our simulation  includes  both UV and X-ray feedback from an accreting BH  and in-situ star formation along with their chemical, mechanical and radiative feedback. Our recipes for Pop III and Pop II star formation are based on \cite{Wise2012} and we insert a MBH of $\rm 10^5~M_{\odot}$ at $z=12$ by turning on its feedback when halo mass reaches well above the atomic cooling limit.  
%We also self-consistently follow transition from Pop III to Pop II stars. 

\begin{itemize}
  \item Our results show that the MBH has accreted only  about 2200 $\rm M_{\odot}$ over the course of  320 Myr and the average mass accretion rate onto the MBH is a few times $\rm 10^{-6}~M_{\odot}/yr$.
  \item The stunted growth of the MBH is a consequence of large outflows driven by SNe  in tandem with AGN feedback which expel  the gas  out of  its surroundings.
 % \item MBH mainly grows during minor mergers and accretion of dense clumps.
  \item Pop III SNe  quickly enrich the ISM with metals in about a few Myr.
  \item The average SFR during the simulation is $ \rm \sim1~M_{\odot}/yr$ and SFR mainly occurs in the primary halo.
 \end{itemize}
 
 The mass accretion rate onto MBH is highly intermittent and peaks around $\rm 10^{-4}~M_{\odot}/yr$.  Interestingly, our results are comparable to the delayed cooling case of high resolution cosmological simulation of \cite{Dubois2015} (hereafter called S15) where  strong SN feedback halts the growth of MBH down to $z=3.5$. Their simulations have effective resolution of 8.7 pc and  halo mass of $\rm 10^{12}~M_{\odot}$ at  $z=2$. They employed simple thermal feedback prescription for  MBH accretion  (for a detailed description see  S15).  In our case energetic SNe from massive stars along with AGN feedback effectively heat the gas  to high temperatures  and yield results similar to the delayed cooling case of  S15. 
%Moreover, dense cold accretion flows and major mergers feed MBH which lead to its rapid growth. Our findings are in agreement with  previous simulations exploring the growth of MBHs  which found that  SN feedback mainly regulates its growth \citep{Habouzit2016b, Prieto2017,DiMatteo2016,Smidt2017}.

Pop III stars quickly enrich the  medium with metals and the first Pop II star cluster  formed 3 Myr after the birth of Pop III stars in our simulation.  The latter have a top heavy initial mass function and thus shorter lives and are initially more effective in removing the gas from the central potential and regulating the growth of  a MBH. The maximum total Pop III stellar mass in our simulation is $\rm \sim 10^5~M_{\odot}$  and the average Pop III stellar mass at any given time is $\rm \sim 10^4~M_{\odot}$.   The Pop II stellar mass becomes greater than Pop III mass after first 5 Myr and  reaches up to $\rm \sim10^9 ~M_{\odot}$ by the end of our simulation. The average SFR is about $\rm \sim 1~M_{\odot}/yr$ and peaks around 10 $\rm M_{\odot}/yr$. The MBH to stellar mass ratio  is $\rm 10^{-3}$  for the first  50 Myr and approaches $\rm 10^{-4}$ in about 100 Myr after the beginning of simulation which is roughly consistent with  the observed BH-stellar mass relation in low mass galaxies in the local universe by \cite{Reines2015}.

%\subsection{Observational Aspects and Caveats}
%The X-ray luminosity of AGN is $\rm \sim 10^{40} ~erg/s$ and an Eddington ratio of $\rm 10^{-3}$ which is much lower compared to the luminous quasars. 
The peak bolometric luminosity from the MBH is about $\rm 4 \times 10^{41}~erg/s$ while the corresponding X-ray luminosity is $\rm  \sim 10^{40}~erg/s$.  This is about two orders of magnitude lower than the  Chandra Deep Field South detection limits for an exposure of 4 Ms. The Eddington ratio is $\rm 10^{-3}$  much lower compared to the luminous quasars. This may explain why such faint sources have not been detected so far. The local analogs of these sources are low-luminosity AGNs provided that the stellar mass-BH relation holds at high redshift \citep{Overzier09,Jia11,Alex2012}. These findings are consistent with current observations of high redshift AGN \citep{Willott2011,Giallongo2015,Weigel2015}.
 We have estimated the flux from the AGN in the near infrared (NIR) band (1-2.2 micron) assuming a bolometric correction of $\rm \sim 7\%$ in the NIR. The estimated flux  is $\rm \sim 10^{-21} ~erg/s/cm^2$ which is about three orders of magnitude lower than the observational limit for the James Webb Space telescope with  NIRCAM  for S/N =10 and integration time of 10 ks.  However,  SFR is about 6 orders of magnitude higher than mass accretion rate onto the MBH.  Therefore it will be difficult to distinguish the AGN from the stellar population \citep{Volonteri2017}. Depending upon the line of the sight, the average column density is $\rm 10^{20}~cm^{-2}$ which is below the Compton thick limit for obscured AGN.  Therefore,  the simulated AGN is not Compton thick but too faint  to be detected with current surveys. This may explain the paucity of AGN in high redshift Lyman Break galaxies.

In our simulation we activated star formation and their feedback at $z=12$ when  halo mass  is  $\rm 10^9~M_{\odot}$. Doing so,  star formation was delayed as Pop III stars are expected to form in $\rm \sim10^8~M_{\odot}$ halos under moderate LW flux which is a shortcoming  of our simulation.  We expect that if Pop III stars were formed in the progenitor halos they would have enriched their hosts and neighboring halos with metals and led to earlier transition from Pop III to Pop II stars.  Also  MBH could not have formed via direct collapse due to early metal enrichment. Such prior star formation may have lowered the gas content in the halo, preheated the gas and consequently the impact of starburst may have been lower. However, we expect that role of such prior Pop III star formation to be  only important during the early phases of galaxy formation.  

The direct collapse BHs are expected to form in metal free halos of  $\rm \sim10^8~M_{\odot}$ at  $ \sim z>10$ but in our simulation we inserted MBH in $\rm 10^9~M_{\odot}$ halo due to the the high computational demand of the radiative transfer calculation. If we had to insert MBH into $\rm \sim10^8~M_{\odot}$ halo, we had to refine even larger region  to resolve the progenitors of the halos. We also assumed here that the MBH formed in one of the progenitor halos  via direct collapse  which mandates that the host halo of  $\rm \sim 10^8~M_{\odot}$ was metal free and  irradiated with a strong LW flux. These conditions should be assessed in future studies to explore their feasibility. In such a scenario, the impact of prior star formation would be less important as it naturally suppresses the star formation in the host halo.  If the MBH was formed via accretion/merging of stellar mass black holes then the feedback from such a BH would have expelled  gas from the halo and regulated the star formation in the host galaxy. Future simulations should self-consistently model  the formation of MBH along with star formation  to better understand their role in galaxy assembly and BH growth in the early universe.

One of the shortcomings of the present work  is  that  cooling is considered only from metals produced by Pop III stars.  Consequently,  the galaxy stellar mass is expected to be lower  by a factor of two compared to the metal cooling both from Pop III and Pop II stars (see \cite{Wise2012a} for a detailed discussion).  Furthermore, we employed the Bondi-Hoyle formalism for accretion onto the MBH which assumes spherical symmetry and is only an approximation for low resolution simulations. In reality, gas around the MBH settles in a rotationally supported disk which cannot be resolved in the present-day numerical simulations. The maximum resolution in our simulation is a few pc which is far from  resolving the accretion disk and flow around the event horizon of a MBH.   Recently \cite{Sugimura2017} performed 2-D radiation hydrodynamical simulations to explore the impact of an-isotropic radiation feedback on  BH growth. They found that  radiation feedback is mainly confined to the bi-polar region while accretion still continues from the equatorial plane. This results in higher accretion rate onto the BH  compared to the an isotropic case and even exceeds the Eddington value.  However,  other studies found that as the disk wind opening angle increases, winds kinematically get coupled with the surrounding gas and lead to an isotropic flow, see \cite{Novak11} and \cite{Ciotti17} for a detailed discussion. It is expected that radiative feedback may become  already isotropic by the time it reaches the  resolution limit (a few parsecs) of our simulation.

Due to the resolution constraints, we employed a Bondi- Hoyle prescription for MBH accretion which does not take into account the angular momentum of the gas  \citep{Debuhr2010,Power2010,Hopkins2011}, and provides only an approximate  solution to the real accretion inflow, which can at times be underestimated or overestimated. When the accretion radius is close to the Bondi radius, however, the Bondi formalism provides a good estimate of the mass flux \citep{Negri2017}.  In future studies, alternative approaches considering the angular momentum transport should be explored. In this study, we only considered the coupling of X-rays with primordial gas chemistry and ignored the interaction of X-rays with metals. The presence of metals can attenuate X-rays and may  impact the formation of HII region as well as the growth of MBH (see \cite{Aykutalp2013}). 
% \ch{Although maximum resolution in our simulation is a few parsecs, but we do not fully resolve the Bondi radius at all times. This may lead to higher star formation rate and consequently larger SNe feedback.  Such strong SNe may even suppress  the mass accretion onto the MBH. So in future studies, it should be ensured that Bondi radius is well resolved to avoid such artifacts.}

 \section*{Acknowledgments}
This work was granted access to the HPC resources of TGCC under the allocation x2016046955 made by GENCI. The research leading to these results has also received funding from the European Research Council under the European Community's Seventh Framework Programme (FP7/2007-2013 Grant Agreement no. 614199, project ``BLACK'').  This project has received funding from the European Union's Horizon 2020 research and innovation programme under the Marie Sklodowska-Curie grant agreement No 656428. JHW is supported by National Science Foundation grants AST-1333360 and AST-1614333, NASA grant NNX17AG23G, and Hubble theory grants HST-AR-13895 and HST-AR-14326.  The simulation results are analyzed using the visualization toolkit for astrophysical data YT  \citep{Turk2011}.
%The research leading to these results has received funding from the European Research Council under the European Community's Seventh Framework Programme (FP7/2007-2013 Grant Agreement no. 614199, project ``BLACK'').
\bibliography{smbhs.bib}
 
\newpage

\end{document}